\documentclass[preprint,eqsecnum,aps,showpacs]{revtex4}
\usepackage{dcolumn}
\usepackage{graphicx}
\usepackage{amsmath}
\usepackage{amssymb}
\usepackage{epsfig}
\usepackage{float}
\usepackage{latexsym}
\usepackage{rotating}
\begin{document}
\title{Kinematics of geodesic flows in stringy black hole backgrounds}
\author{
Anirvan Dasgupta\footnote{Electronic address: {\em anir@mech.iitkgp.ernet.in}}${}^{}$}
\affiliation{Department of Mechanical Engineering and Centre for Theoretical Studies  \\
Indian Institute of Technology, Kharagpur 721 302, India}
\author{
Hemwati Nandan\footnote{Electronic address: {\em hnandan@cts.iitkgp.ernet.in}}${}^{}$}
\affiliation{Centre for Theoretical Studies  \\
Indian Institute of Technology, Kharagpur 721 302, India}
\author{
Sayan Kar\footnote{Electronic address: {\em sayan@cts.iitkgp.ernet.in}}${}^{}$}
\affiliation{Department of Physics and Centre for Theoretical Studies  \\
Indian Institute of Technology, Kharagpur 721 302, India}
\begin{abstract}

We study the kinematics of timelike geodesic congruences in two and four
dimensions in spacetime geometries representing  stringy black holes.
The Raychaudhuri equations for the kinematical quantities 
(namely, expansion, shear and rotation) characterising such geodesic
flows are written down and
subsequently  solved analytically (in two dimensions) and numerically (in four
dimensions)  for specific geodesics flows. 
We compare between geodesic flows in dual (electric and magnetic) stringy black hole backgrounds in four dimensions, 
by showing the differences that arise in the corresponding evolutions 
of the kinematic variables. 
The crucial role of initial conditions and
the spacetime curvature on the 
evolution of the kinematical variables is illustrated. Some novel general conclusions
on caustic formation and geodesic focusing are obtained from the analytical and numerical findings.
We also propose a new quantifier in terms of the time (affine parameter) of approach to a 
singularity,
% and (b) the location of extrema in the functional evolution of the kinematic variables, 
which may be used to distinguish between flows in different geometries.
In summary, our quantitative findings bring out hitherto unknown features of the kinematics of geodesic flows, which, otherwise, would have remained overlooked, if we confined ourselves to only a qualitative analysis.
\end{abstract}

%\noindent{Keywords}: Geodesics, Raychaudhuri equations, stringy black hole.

%04.25.Dm (Numerical relativity)
%04.20.Cv (Fundamental problems and general formalism)
%97.60.Lf(black holes)
\maketitle

\section{Introduction}
The kinematics of geodesic congruences is
characterised by three kinematical quantities: isotropic expansion, shear and rotation
(henceforth referred as ESR) \cite{toolkit,wald,joshi,ellis,ciufolani,review}.
The evolution of these quantities along the geodesic flow,
are obtained from the Raychaudhuri equations \cite{toolkit,wald,joshi,ellis,ciufolani,review}.
These equations are derived by relating the evolution of the deformation
(or deviation) vector
between two neighbouring geodesics (expressed in terms of the ESR variables) to the curvature of the
space/spacetime.
%is considered to study the deformations in a medium.
This is already a well-studied subject (see \cite{review} and the references therein).

In its original incarnation, the Raychaudhuri equation provided the basis for 
the description and analysis of
spacetime singularities in gravitation and cosmology \cite{hawk}. For example,
the equation for the expansion  and resulting theorem on geodesic focusing is 
a crucial ingredient in the proofs of
Penrose-Hawking singularity theorems \cite{Penro,Haw}. However, these equations have a much wider scope in studying geodesic as well as
non-geodesic flows in nature which may possibly arise in diverse contexts (see \cite{review} for some open issues).
We have recently used these equations to investigate the kinematics of flows on
flat and curved deformable media (including elastic
and viscoelastic media) in detail \cite{adg1,adg}. 

In this article, we attempt to understand
the kinematics
of geodesic flows in the presence of spacetime geometries representing black holes. Though, it is true that the Raychaudhuri equations have been around now
for more than half a century, we are not aware of any attempt at
a complete study of its solutions and the dependence of the ESR variables
on the initial conditions imposed on them (however, see \cite{prd} for a 
recent work). A slight subtlety may be noted here.
The Raychaudhuri equations are for the ESR variables, but they also involve the tangent
vector field (denoted as $u^i$, later). Thus, unless one knows the solutions for the $u^i$ (i.e. the first integrals of the geodesic equations) 
one cannot proceed towards solving for the ESR. 
Alternatively, one can try to find solutions
for the full set of variables, i.e. $u^i$ as well as the ESR, 
by imposing initial
conditions on all of them and evolving the full system of equations
along the flow. 
This approach enables us to obtain the tangent vector field
as well as the ESR simultaneously. In our work, we adopt this method
primarily for the four dimensional cases 
where we are unable to solve for the geodesics or the
$u^i$ analytically.
  
As an aside, it may be noted, that the study of accretion of matter, or 
evolution of spacetime deformations near a black hole has been
a fascinating topic of study in classical general relativity. 
Such processes can perhaps be investigated
using the Raychaudhuri equations. The study of kinematics of non-spacelike congruences may further provide useful insights on the  observable features in a given spacetime. We take this as a background motivation for 
our study.

The non-trivial geometry of different types of  black holes is of crucial interest in general
relativity and for our purpose, we will consider some typical
two and four dimensional stringy black hole spacetimes. 
The two dimensional black hole geometry we work with was first obtained in the context of
string theory \cite{Witten}-\cite{harvey} in the nineties by Mandal et. al. \cite{Mandal}.
Exact solutions for geodesics and geodesic deviation in this two dimensional 
stringy black hole background have already
been studied earlier \cite{rk}. In four dimensions, the simplest eternal black hole
geometry is, of course, the Schwarzschild. Rather than working with
 just the Schwarzschild alone, we consider the variations of the Schwarzschild which
have arisen in the context of string theory \cite{ghs,gthorowitz,gsw}. All the four dimensional 
geometries we choose to work with have a Schwarzschild limit (obtainable
by setting a parameter in the line element to zero). Further, 
the validity of the
various (weak, strong, null averaged) energy conditions
for the matter that threads such
stringy black hole spacetimes have also been investigated in detail in order to understand
the nature of geodesic focusing \cite{skprd97}.\\

Our article is organised as follows.
We first review the background spacetimes (Section II) 
and then (Section III) derive the evolution (geodesic and Raychaudhuri)
equations for the  two dimensional stringy black hole metric. In this 2D case, 
the Raychaudhuri equation is just the equation for the expansion scalar. 
We obtain analytical solutions for the
expansion scalar, based on
which the role of initial conditions on caustic formation and 
geodesic focusing/defocusing is 
investigated and analysed.
Subsequently (Section IV), we turn to actual four dimensional solutions in dilaton-Maxwell gravity. We analyse
the nature of deformations for the geodesic flows in the Garfinkle-Horowitz-Strominger
(GHS) electric and magnetic (dual) solutions \cite{ghs} and compare our results with those for
Schwarzschild geometry
(obtainable from the GHS metrics by setting a stringy parameter to zero).  Here, the kinematics of deformations is
studied on the two
dimensional equatorial plane. In the absence of analytical solutions, we solve the geodesic
and Raychaudhuri equations numerically under differing initial 
conditions on the associated variables.
The generic features of the evolution of deformations are brought out. The influence of the
gravitational field  on the evolution of the ESR is
discussed and the effect of curvature is understood. Finally (Section V), we conclude  our results and suggest some relevant future work.

\section{The stringy black hole spacetimes} \label{4dsbhst}
In this section, we quickly recall a well-known solution obtained in the context of 
two dimensional low energy effective string theory {\cite{gsw}}. The
line element which we mention below solves the so-called $\beta$-function
equations for the string $\sigma$-model {\cite{gsw}}.  It is known that with appropriate methods
of compactification one can obtain effective equations in two as well as
other higher dimensions.
These equations, in the simplest scenario, are for the metric field and the dilaton field. 
We quote below the line element in two dimensions \cite{Mandal,Witten},
\begin{equation}
ds^2 = - (1-\frac{m}{r})\,  dt^2 + \frac{\kappa \, dr^2}{ 4 r^2\,  (1-\frac{m}{r})};\, \,  \,\, \, \, \, \, \, (r \geq m),
\label{metric2d}
\end{equation}
where, $m$ and $\kappa$ (with dimensions of length square) are the mass and central 
charge parameters, respectively, and are linked with the concepts in two dimensional 
string theory \cite{skprd97} which we do not bother about here.\\
In $3+1$ dimensions, we also have asymptotically flat solutions representing black holes 
in dilaton-Maxwell gravity. Such solutions, due to Garfinkle, Horowitz and Strominger \cite{ghs},
represent electric and dual magnetic black holes
\cite{gthorowitz}. The spacetime geometry of these line elements are
 causally similar to Schwarzschild geometry. The metric for the black 
 hole with electric charge is given as,
\begin{equation}
ds^2 = - \frac{(1-\frac{m}{r})}{ \left(1+ \frac{m \, \sinh^2{\alpha}}{r}\right)^{2}}\, dt^2 + \frac{dr^2}{ (1-\frac{m}{r})} \,  +  r^2 \,  d\Omega_2^2
\label{metric4da},
\end{equation}
where  $d\Omega_2^2 = (d \psi^2 + \sin^2\psi\, d\phi^2)$ is the  metric on a two dimensional unit sphere and $\alpha$ is a parameter related to the electric charge.  Further, 
the dual (magnetic) metric of (\ref{metric4da}) is given as follows \cite{gthorowitz}, 
\begin{equation}
ds^2 =  - \frac{(1-\frac{m}{r})}{ (1- \frac{Q^2}{m r})}\, dt^2 + \frac{dr^2}{(1-\frac{m}{r})\,(1- \frac{Q^2}{m r})} \,  +  r^2 \,  d{\Omega_2}^2
\label{metric4dq}.
\end{equation}
where $Q$ is the magnetic charge of the black hole. In the respective limit of $\alpha =0$ or
$Q=0$, we have Schwarzschild geometry in both the cases. 
We shall investigate the ESR variables for geodesic flows in each of the above
two four dimensional metrics and compare our results for the electric
and magnetic solutions with those for the Schwarzschild.\\ 
For a general and compact representation of the above spacetimes, one can use a generic line element in $3+1$ dimensions as follows,
\begin{equation}
ds^2 = - X(r)\, dt^2 +  Y(r)\,{dr^2} \,  +  r^2 \,  d\Omega_2^2
\label{gmetric},
\end{equation}
 for specific choices of $X(r)$ and $Y(r)$. The general structure of the geodesic equations for the line element (\ref{gmetric}) are given by,
 \begin{equation}
\ddot t  + \frac{ X'(r)} {X(r)}\, \dot r \, \dot t  = 0,
\label{tg11}
\end{equation}
\begin{equation}
\ddot r  + \left (\frac{ X'(r) \, \dot t^2 +  Y'(r) \,\dot r^2  - 2 r \, \dot \psi^2 -2 r \, \sin^2 \psi \, \dot \phi^2} {2 \, Y (r)} \right ) = 0,
\label{rg11}
\end{equation}
\begin{equation}
\ddot \psi + \frac{2}{r}\, \dot r \, \dot \psi - \cos \psi \, \sin \psi\,  \dot \phi^2  = 0,
\label{ps1}
\end{equation} 
\begin{equation}
\ddot \phi + \frac{2}{r} \, \dot r \, \dot \phi + 2 \, \cot \psi \, \dot \psi \,  \dot \phi =0,
\label{ph1}
\end{equation}
where the prime denotes the differentiation with respect to $r$. Using the 
functional forms of $X(r)$ and $Y(r)$ 
(corresponding to the line elements given by (\ref{metric4da}) and (\ref{metric4dq}), in equations (\ref{tg11}) and (\ref{rg11}),  
one can easily obtain the set of geodesic equations 
for the particular cases of the electric and magnetic stringy black holes.  For  the line element (\ref{metric2d}) in $1+1$ dimensions, the term $r^2 \,  d\Omega_2^2$ is absent in (\ref{gmetric}) and the geodesic equations are given by (\ref{tg11}) and (\ref{rg11}) (without the terms $\psi$ and $\phi$).

\section{Kinematics of deformations in $1+1$ dimensions} \label{2dcase}

\subsection{Kinematic variables}

The evolution of space-like deformations in a two dimensional geodesic 
congruence is captured through  the evolution of the geodesic deviation 
vector $\xi^{i}$ (where $i=1,2$) on a space-like hypersurface. These 
deformations can be described in terms of a second rank tensor 
$B^{i}_{j}= \nabla_j u^i$ \cite{toolkit,adg1}, which governs the dynamics of the congruence. The second order derivative of  the  vector ${\xi}^i$ with respect to an affine parameter is given as follows \cite{adg,adg1},
\begin{equation}
{\ddot\xi}^i = (\dot B^i_{j} + B^i_{\,\,k}B^k_{\,\,j})\, \,  \xi^j \,.
\label{ddotxi}
\end{equation}
The evolution tensor $B^i_{j}$ in equation (\ref{ddotxi}) is usually decomposed into irreducible parts signifying the expansion (scalar $\theta$), shear (trace-free tensor $\sigma^i_{j} $) and  rotation (antisymmetric tensor $\omega^i_{j} $). Since  $u_i\sigma^i_{j} = 0$ (space-like deformations), and using the zero trace property, i.e., $\sigma^i_{\,i} =0$, leads to  $\sigma^{0}_{\,0}   = \sigma^{1}_{\,1} = \sigma^{1}_{\,0} =0$. Similarly, the rotation tensor $\omega^{i}_{\,j}$  also satisfies  $u_i\omega^{i}_{\,j} = 0$ which leads to  $\omega^{1}_{\,0} = \omega^{0}_{\,1} = 0$.  Therefore, the evolution tensor can be expressed 
only in terms of the expansion scalar, in the following form,
\begin{equation}
B^i_{j} = \theta \, h^i_{\, j},
\label{bth}
\end{equation}
where the projection metric is defined as $h^{i}_{\,j} = \delta^{i}_{\,j} + u^i\, u_j$. Here,  $u_i$ is a time-like vector field.
\subsection{The evolution equations} The evolution equations for a congruence of  time-like geodesics for the present case consist of the Raychaudhuri equation for the expansion scalar and the geodesic equations derived for a particular metric. In  order to derive the Raychaudhuri equation for the expansion scalar, we first write down the second derivative of the  deformation vector in the following form, 
\begin{equation}
\ddot \xi^{i}\, = - R^{i} _{\,\, ljm} u^{l}u^{m}\xi^{j} . \\
\label{sdform}
\end{equation}

%The equations (\ref{ddotxi}) and (\ref{sdform}) then lead to the time evaluation of ${B}^i_{\,\,j}$ as follows,
%\begin{equation}
%\dot{B}^i_{\,\,j}+B^{i}_{\,\,k}B^k_{\,\,j} + K^i_{\,\,j}+\beta B^i_{\,\,j}= -  R^{i} _{\,\, ljm} u^{l}u^{m},
%\label{dotb}
%\end{equation}
%which consequently leads to  the evolution (Raychaudhuri) equations for the ESR variables.
\subsubsection{Raychaudhuri equation for expansion scalar}
%In order to calculate the Raychaudhuri equation of expansion and geodesic equations, let us consider a  two dimensional stringy black hole spacetime with the metric  given in the following form \cite{rk},
%\begin{equation}
%ds^2 = - (1-\frac{m}{r})\,  dt^2 + \frac{\kappa \, dr^2}{ 4 r^2\,  (1-\frac{m}{r})} \,,
%\label{metric2d}
%\end{equation}
%where, $m$ and $\kappa$ with the dimensions of length square are the mass and central charge parameters, respectively, which are described from the perspectives of two dimensional string theory \cite{skprd97}.(\ref{ph1})
 
Using the equation (\ref{bth}) and (\ref{sdform}) in the equation (\ref{ddotxi}), one can now obtain the Raychaudhuri equation for the expansion scalar as given below,
\begin{equation}
\dot \theta + {\theta^2} = - R^{i} _{\,\, lim} u^{l}u^{m} = - R_{\,\, lm} u^{l}u^{m}.
\label{theta}
\end{equation}
%\begin{equation}
%\dot \sigma_{+} \, + \, (\beta +\theta) \, \sigma_{+}  +  k_{+}  \, + \,  \frac{m \, \dot r^2} {2 r^2\,  (r-m)} - \frac{m}{\kappa r}  =0,
%\label{sigma+}
%\end{equation}
%\begin{equation}
%\dot \sigma_{\times} \, + (\beta +\theta) \, \sigma_{\times} \, +  k_{\times} \, + \frac{m}{r^2} \,\left[\frac{r-m}{\kappa} -\frac{1}{4(r-m)} \right ]\,  \, \dot r \, \dot t  =0,
%\label{sigmacross0}
%\end{equation}
%\begin{equation}
%\dot \omega+ (\beta + \theta)\, \omega  +  \frac{m}{r^2} \,\left[\frac{r-m}{\kappa} + \frac{1}{4(r-m)} \right ]\, \,   \dot r \,\, \dot t =0
%.\label{omega}
%\end{equation}
The general form of equation  (\ref{theta}) for the metric  (\ref{metric2d}) can also be written as follows:
\begin{equation}
\dot \theta + {\theta^2} - \frac{R}{2} = 0,
\label{theta1}
\end{equation}
where the Ricci scalar $ R= 4 m / \kappa r$. It may be noted that for $r \rightarrow \infty$, the  equation (\ref{theta1}) reduces to that in flat space without shear and vorticity. This evolution equation (\ref{theta1}) along with the geodesic equations corresponding to the metric (\ref{metric2d}), form a complete set of equations required for studying the kinematics of deformations of geodesic congruences. 

\subsubsection{First integrals of geodesic equations}
The first integrals of the geodesic equations in $1+1$ dimensions can be obtained as,
\begin{equation}
\dot t = \frac{E}{2 \, (1-\frac{m}{r})} ,
\label{tg1}
\end{equation}
\begin{equation}
\dot r^2 =  \frac{1}{\kappa} \,\, [ \, (E^2-4) \, r^2 + 4 m r\, ] = -V_2(r), \label{rdot}
\end{equation}
%\begin{equation}
%\dot r^2 =  \frac{1}{\kappa} \,\, [ \, (E^2-4) \, r^2 + 4 m r\, ]. \label{rdot}\end{equation}
where  $V_2 (r)$ is an effective potential. It may be noted that the constraint  $g_{\alpha \beta} u^{\alpha} u^{\beta} = -1$ (time-like geodesics) is used to obtain (\ref{rdot}).
The different choices for the constant of motion $E$ result in the different behaviour of the effective potential. One can have a harmonic, or an inverted harmonic oscillator corresponding to $E^2 < 4$, or $E^2 > 4$, respectively, while $E^2=4$ results in a linear potential with negative slope \cite{rk}. We will now solve the Raychaudhuri equation for the expansion scalar for all the choices of $E$ mentioned above.

\subsection{Exact solution for expansion scalar} \label{eslam}
The equation (\ref{theta1}) can be solved for the above-mentioned three different cases by integrating  equation (\ref{rdot}) once  (see \cite{rk}), and then using $r$  in equation (\ref{theta1}).  \\
{{\it Case (A)} : $E^2 < 4$.}\\
The equation (\ref{theta1}) for  the expansion scalar with $E^2 < 4$ reads:
\begin{equation}
\dot{\bar\theta}(\bar \lambda) + \bar{\theta}^2 (\bar \lambda) - 2 \sec^2 \bar \lambda =0,\label{theta2}
\end{equation}
where we have used the scaling $\{\bar \lambda,\bar\theta\}= [(4-E^2)/4\kappa]^{1/2}\{\lambda,\theta\}$. 
The solution of equation (\ref{theta2}) is then given by,
\begin{equation}
\bar\theta(\bar \lambda) =  \frac{\tan {\bar \lambda} + (D_1 + {\bar \lambda}) \sec^2{\bar \lambda}}{1+ (D_1+ {\bar \lambda}) \tan {\bar \lambda}},
 \label{ts1}
\end{equation}
where $D_1$ is an integration constant which can be given in terms of the initial conditions as follows,
\begin{equation}
D_1 =  \frac{(1- {\bar \lambda}_0 \bar\theta_0 ) \tan {\bar \lambda}_0 + {\bar \lambda}_0 \sec^2{\bar \lambda}_0 -\bar\theta_0}{\bar\theta_0 \tan {\bar \lambda}_0 -\sec^2{\bar \lambda}_0 }.
 \label{d10}
\end{equation}

\noindent{{\it Case (B)} : $E^2 = 4$.}\\
 The equation for expansion scalar with $E^2 = 4$ reads:
\begin{equation}
\dot \theta (\lambda) + {\theta ^2(\lambda)} - \frac{2}{\lambda^2} = 0,
\label{theta3}
\end{equation}
and the solution of equation (\ref{theta3}) can be given as follows,
\begin{equation}
\theta(\lambda) =  \frac{2 D_2 \lambda^3 -1}{\lambda(D_2 \lambda^3+1)}. \label{ts2} \end{equation}
where the integration constant $D_2$  is given as,
\begin{equation}
D_2 =  -\frac{(1+ {\lambda}_0 \theta_0)}{{\lambda}_0^{3}  ({\lambda}_0 \theta_0 -2)}.
 \label{d20}
\end{equation}

\noindent{{\it Case (C)} : $E^2 > 4$.}\\
The equation for the expansion scalar for this case reads:\\
\begin{equation}
\dot{\bar\theta}(\bar \lambda) + \bar{\theta}^2(\bar \lambda) - 2 \, {\rm cosech}^2{\bar \lambda}=0\label{theta4},
\end{equation}
where $\{\bar \lambda,\bar\theta\}= [(E^2-4)/4\kappa]^{1/2}\{\lambda,\theta\}$. 
The solution of equation (\ref{theta4}) is,
\begin{equation}
\bar\theta(\bar \lambda) =  \frac{\coth {\bar \lambda} -(D_3 + {\bar \lambda}) \, {\rm cosech}^2 {\bar \lambda}} {(D_3 + {\bar \lambda}) \coth {\bar \lambda} -1} \label{ts3}.\end{equation}
where the integration constant $D_3$ is given as,
 \begin{equation}
D_3 =  \frac{(1- {\bar \lambda}_0 \bar\theta_0) \coth {\bar \lambda}_0 - {\bar \lambda}_0 \, {\rm cosech}^2{\bar \lambda}_0 +\bar\theta_0}{\bar\theta_0 \coth {\bar \lambda}_0 + {\rm cosech}^2{\bar \lambda}_0 }.
 \label{d30}\end{equation}

%\subsection{The expansion scalar as a function of coordinates}
It is in order to mention here that, following an altogether different approach, one can calculate the 
expansion scalar $\theta$ as a function of $r$ from the expressions of the first integrals 
$u^i=(\dot t$, $\dot r)$ (the velocity field)
given by (\ref{tg1})-(\ref{rdot}) (i.e., without integrating the Raychaudhuri equation). 
Using $\theta=\nabla_i u^i$, we have
\begin{equation}
\theta =-\frac{2m}{\kappa\dot{r}}=\mp \frac{ 2m}{\sqrt{\kappa[(E^2-4)r^2 + 4mr]}}.
\label{revth}
\end{equation}
In view of the solutions of $\theta$ obtained earlier in this section by integrating the 
Raychaudhuri equation, the expression in (\ref{revth}) deserves attention. 
This expression for $\theta$ shows that a caustic forms at a turning point of the geodesic
motion (i.e., where $\dot r=0$). For $\dot{r}\rightarrow 0^+$ ($\dot{r}\rightarrow 0^-$),
we have focusing (defocusing).
%One may note that the sign of $\theta$ is opposite to that of $\dot r$. 
In order to obtain the explicit $\lambda$ dependence of $\theta$, one has to substitute $r(\lambda)$ from the
solution of the geodesic equation for $r$ in (\ref{revth}). It may be easily checked that the solutions
thus obtained from (\ref{revth}) are special cases of the previous solutions.
%and finally use the initial condition $\theta(\lambda_0)=\theta_0$. 
It is important to note that, unlike the solutions of $\theta$ obtained by integrating the
Raychaudhuri equation, in the expression (\ref{revth}), there is no way of specifying any initial
condition on $\theta$. Thus, using (\ref{revth}), one cannot study the effect of initial conditions 
on the evolution of a geodesic congruence. This is a subtle issue which will be discussed further in the 
following section.

%Alternatively, one may substitute $r(\lambda)$ in the Raychaudhuri equation
%and integrate it to obtain $\theta(\lambda)$ directly. We follow this
%approach below, though we mention that the results are the same, either way.  

\subsection{Analysis of geodesic focusing}
From the exact solutions of the expansion scalar obtained above by integrating the Raychauduri equations, 
one can determine the occurrence of finite time singularity (i.e., caustic formation). For Case (A), 
it may be deduced from (\ref{ts1}) that
the expansion scalar $\bar\theta\rightarrow \pm\infty$ as 
$\bar{\lambda}\rightarrow\pi/2$. Thus, we may have focusing or
defocusing of geodesic congruences depending on the initial conditions.
One may therefore calculate a critical initial value of the expansion scalar by exploiting the indefiniteness 
condition on $\theta$ which leads to $D_1= -\pi /2$ ( for ${\bar \lambda}_0 < \pi / 2$). 
Now from (\ref{d10}), one can calculate the critical initial value of the expansion scalar as,
\begin{equation}
\bar\theta_0^{c} =  \frac{(\frac{\pi}{2} - {\bar \lambda}_0) \sec^2{\bar \lambda}_0 -  \tan {\bar \lambda}_0}{(\frac{\pi}{2} - {\bar \lambda}_0) \tan {\bar \lambda}_0-1},\qquad ({\bar \lambda}_0 < \frac{\pi}{2}).
 \label{tc0}
\end{equation}   
 From this analysis and using ${\bar \lambda}_0 =1$, we have $\bar\theta_0^c \sim-1.95$. Thus, for $\bar\theta_0 < \bar\theta_0^c$, we have the congruence focusing, i.e., a finite time singularity occurs.

In Case (B), it may easily be concluded that caustic in the geodesic congruence forms for any
initial condition $\theta_0<-1/\lambda_0$. In this case, we have geodesic focusing whenever the
initial condition satisfies this condition.

For Case (C), it may be observed from the solution (\ref{theta4}) that defocusing is not possible. However, focusing  can occur 
for an appropriate choice of initial conditions  which can be obtained by choosing $\bar\theta_0<\bar\theta_0^{c}$ 
where $\bar\theta_0^{c} = -2/ \sinh 2{\bar \lambda}_0$ 
is the critical value of the initial expansion scalar. For  $\bar\theta_0>\bar\theta_0^{c}$, finite time singularity cannot occur.

It is well-known from the work of Tipler \cite{tipler} that if $R_{lm} u^l u^m \geq 0$ (timelike
convergence condition) then
focusing (and conjugate points) arise in the congruence within a finite value of the affine parameter. 
From our above analysis, it may thus  seem counter--intuitive that $R_{lm}u^l u^m = -R/2 \leq 0$ leads to focusing of 
timelike 
geodesic congruences in two dimensional spacetimes. This, however, is not in conflict with 
the results of Tipler. 
In situations, such as those presented above, the timelike convergence condition is clearly violated.  But, with
appropriate initial conditions on the
expansion (as shown above), one may still have a focusing of geodesic congruences. Therefore, we may say that
initial conditions have a crucial role to play in focusing. 
%We shall demonstrate more of this aspect when we discuss the four dimensional cases below. 

The Case (C) (i.e., with $E^2>4$) brings out a subtle and interesting difference between the solutions of $\theta$
in (\ref{theta4}) and (\ref{revth}). As is clear from the expression of the effective potential in (\ref{rdot}),
all outward trajectories (outside the horizon) escape out to infinity without any turning point. 
Hence, one may conclude from (\ref{revth}) that there are no caustics in such a scenario.
However, from the solution (\ref{theta4}) discussed above, we do have focusing depending on the
initial condition $\theta_0$. This difference can be reconciled with if we realize that the 
expression (\ref{revth}) actually yields the expansion scalar
field corresponding to the (static) velocity vector field $u^i=(\dot{t},\;\dot{r})$. On the other hand, the
expression (\ref{theta4}) tells us about the expansion history of a congruence, which may have been started
with an arbitrary initial expansion, as observed in the local frame of a freely falling observer. 
This is also the approach adopted while proving the well-known focusing theorem \cite{toolkit,wald,joshi}.
  
\section{Kinematics of Deformations in $3+1$ Dimensions}
\subsection{The evolution equations}
In four dimensions, for a congruence of time-like geodesics, the transverse 
metric on a space-like hypersurface can be expressed as,
\begin{equation}
h_{\alpha \beta}=  g_{\alpha \beta}+ u_{\alpha} u_{\beta},\, \, \, \, \, \, \, \, \, \, \, \, (\alpha, \beta = 0,1,2,3),\label{gab}
\end{equation}
where $u^\alpha$ is the time-like vector field tangent to the geodesic at each point satisfying  $u_{\alpha} u^{\alpha} =-1$, and the four dimensional metric $g_{\alpha \beta}$ is defined through the line elements (\ref{metric4da}) and (\ref{metric4dq}). The transverse metric satisfies $u^\alpha h_{\alpha \beta} = 0$, i.e., $h_{\alpha \beta}$ is orthogonal to $u^{\alpha}$. This transverse space-like hypersurface represents the local rest frame of a freely falling observer in the given spacetime. The point of interest in this investigation is to determine the deformations in this local rest frame as perceived  by the observer. The evolution of space-like deformations on this transverse hypersurface can be quantified using the tensor $B_{\alpha \beta}$, which  can now be decomposed as follows,
\begin{equation}
B_{\alpha \beta} = \frac{1}{3}{\theta} \, h_{\alpha \beta} +\sigma_{\alpha \beta} + \omega_{\alpha \beta}, \label{bij3d}
\end{equation}
where $\theta = B^{\alpha}_{\, \, \alpha}$ is the expansion scalar, while $\sigma_{\alpha \beta} = B_{(\alpha \beta)} - \theta \, h_{\alpha \beta} /3 $  and $ \omega_{\alpha \beta} = B_{[\alpha \beta ]}$ are the shear and rotation tensors.
The brackets $( \, )$ and $[\, ]$ denote symmetrisation and antisymmetrisation, respectively. The shear and rotation tensors also satisfy  $ h^{\alpha \beta}\, \sigma_{\alpha \beta} = 0 $ and $ h^{ \alpha \beta}\, \omega_{\alpha \beta} = 0$, as can be easily checked. The evolution equation for $B_{\alpha \beta}$  takes the form,
\begin{equation}
\dot{B}_{\alpha \beta} + B_{\alpha \gamma} B^\gamma_{\,\,\beta}  = -  R_{\alpha \eta \beta \delta} \, \, u^{\eta}u^{\delta}.
\label{dotb1}
\end{equation}
Using this equation, one can now obtain the evolution equations for the ESR variables 
which are discussed below.

%HERE
\subsubsection{Raychaudhuri equations}
 Following well--known methods, the Raychaudhuri equations for the expansion scalar, and the shear and rotation tensors 
can be obtained as,
\begin{equation}
\dot \theta + \frac{1}{3}{\theta^2} + (\sigma^2 -\omega^2) + R_{\alpha \beta} \, u^{\alpha} \, u^{\beta} = 0,
\label{theta3e}
\end{equation}
\begin{equation}
\dot \sigma_{\alpha \beta} + \frac{2}{3} \theta \, \sigma_ {\alpha \beta} +  \sigma_ {\alpha \gamma} \sigma^{\gamma}_{\, \, \beta} +  \omega_ {\alpha \gamma} \omega^{\gamma}_{\, \, \beta} -   \frac{1}{3} (\sigma^2 -\omega^2) \, h_{\alpha \beta} + C_{\alpha \eta \beta \delta}\, u^{\eta} \, u^{\delta}   - \frac{1}{2} \tilde R_{\alpha \beta} = 0,
\label{sigma3}
\end{equation}
\begin{equation}
\dot \omega _{\alpha \beta} + \frac{2}{3} \theta \, \omega_{\alpha \beta} + \sigma_{\alpha}^{\, \, \gamma} \, \omega_{\gamma \beta} + \omega_{\alpha}^{\, \, \gamma} \, \sigma_{ \gamma \beta} = 0, \label{omega3}
\end{equation}
where $\sigma^2 = \sigma_ {\alpha \beta} \,\sigma^{\alpha \beta}$,  $\omega^2 = \omega_ {\alpha \beta} \,\omega^{\alpha \beta}$, $\tilde R_ {\alpha \beta} = h_{\alpha \gamma} \, h_{\beta \delta} R ^{\gamma \delta} - h_{\alpha \beta} \, h_{\gamma \delta} R ^{\gamma \delta} / 3$ and  $C_{\alpha \beta \eta \delta}$ is the Weyl tensor.  These equations are first-order, coupled, nonlinear and inhomogeneous differential equations. The equation (\ref{theta3e}) for the 
expansion is the well-known Riccati equation, and, as mentioned before,
is of prime importance in the context of the proof of the singularity theorems in general relativity \cite{Penro,Haw} and in establishing the notion of geodesic focusing \cite{review}. With the projection metric defined in (\ref{gab}), the Raychaudhuri equations essentially turn out to be structurally similar as in the case of three spatial dimensions. It may also be noted that there can be some congruences having a vanishing vorticity  for which the velocity vector field  is hypersurface orthogonal, and (\ref{omega3}) becomes identically zero. 

\subsubsection{First integrals of geodesic equations on the equatorial section} \label{gev}

One may note that the equations (\ref{ps1}) and 
(\ref{ph1}) are independent of $X(r)$, $Y(r)$ and their derivatives.
 Without the loss of generality, one can choose $\psi = \pi/2$ which 
satisfies (\ref{ps1}) identically. With this choice, we can 
capture the kinematics of
deformations in the $r$-$\phi$ plane. One can now integrate (\ref{ph1}) once to obtain $\dot \phi = C/r^2$ where $C$ is a constant of motion. We will hereafter use these considerations. It is also noteworthy that with $\psi=\pi /2$, this four dimensional description reduces to a three dimensional one and the Raychaudhuri equations for the components of shear and rotation can be calculated in a 
way similar to our recent work (see the reference \cite{adg1}). In addition, 
we must keep in mind that
the time-like vector field $u^i$ satisfies the normalisation condition
$u^i u_i =-1$, which leads to,
\begin{equation}
r^2 \, [\,  - X(r) \, \dot t^2  +  Y(r) \,\dot r^2  \,  + 1] +  C^2   =0.
\label{cons1}
\end{equation}
The above constraint (\ref{cons1}) also represents a first integral of the set 
of geodesic equations (\ref{tg11})-(\ref{ph1}) for a specific choice of the 
constant of integration. We will now discuss the geodesic equations and effective potentials for the cases corresponding to the line elements (\ref{metric4da}) and  (\ref{metric4dq}) respectively.\\

\noindent {\bf Case I } (corresponding to line element (\ref{metric4da})) \\

\noindent The first integral of equation (\ref{tg11}) for this case is calculated as follows,
\begin{equation}
\dot t = \frac{E \, (r+ m \, \sinh^2\alpha)^2}{2 \,r\, (r-m)}.
\label{tg1a}
\end{equation} 
Now, using (\ref{tg1a}) in the constraint (\ref{cons1}) leads to,
\begin{equation}
\dot r ^2 =  \frac{1}{4\, r^2} \,\, [ E^2 \, (r+ m \sinh^2\alpha)^2  + 4 m r\, - 4 r^2 + \frac{4 C^2} {r} (m-r)\,]  = - V_4^E \, (r) 
\label{potentialv4a}, 
\end{equation}
where $V_4^E(r)$ is the effective potential for the case of electric black hole.
The effect of the parameter $\alpha$ on the radial motion for different values of $E$ and $C$ can be visualised directly from (\ref{potentialv4a}). The orbits (circular, scattering and plunge) appear to be qualitatively 
similar to those in Schwarzschild geometry (see \cite{Hartle}).\\
  
\noindent {\bf Case II } (corresponding to line element (\ref{metric4dq}))\\

\noindent The first integral of equation (\ref{tg11}) for the magnetic case is 
as follows,
\begin{equation}
\dot t = \frac{E\,  (m r- {Q^2})}{2 \, m \, (r-m)}.
\label{tg1q}
\end{equation}
The constraint (\ref{cons1}) along with equation (\ref{tg1q}) then leads to,
\begin{equation}
\dot r ^2 =  \frac{1}{4\, r^2} \,\, [ E^2\,r^2 +  \frac{mr}{(m r-{Q^2})} \, ( 4 m r\, - 4 r^2 + \frac{4 C^2} {r} (m-r) \, )\,]  = - V_4^M \, (r) 
\label{potentialv4q}, 
\end{equation}
where $V_4^{M}(r)$ is the effective potential for the magnetic black hole.
The structure of orbits are qualitatively same as in Case I.\\

\noindent The causal structure of the electric and magnetic
stringy black hole spacetimes is similar to the Schwarzschild geometry \cite{gthorowitz}. This provides a motivation for comparing these three cases. The Schwarzschild metric can be  constructed from (\ref{gmetric}) with, $X(r) = (1-m/r)$ and $ Y(r) =(1-m/r)^{-1}$ where usually $m=2M$ with $M$ as the mass of the Schwarzschild black hole. Later, we will consider  $m=1$ for numerical computations. The Raychaudhuri equations in the Schwarzschild case follow from (\ref{theta3e})-(\ref{omega3}) with $ R_{\alpha \beta} \, u^{\alpha} \, u^{\beta}=0$  and $ \tilde R_{\alpha \beta} =0$. The geodesic equations are well-known and the first integrals of the $t$ and $\phi$ equations are same as those for the stringy black holes in $3+1$ dimensions (see Section \ref{gev}) with $\psi=\pi/2$. The first integral of the $r$ equation which satisfies the time-like constraint (\ref{cons1}) on the velocity field leads to the following effective potential,
\begin{equation}
V_4^{S}(r) = - \frac{1}{4r^2}\left[(E^2-4)r^2 + 4 m r+\frac{4 C^2}{r}(m -r)\right]. \label{potsc}
\end{equation}
\begin{figure}
\centerline
\centerline{\includegraphics[scale=0.58]{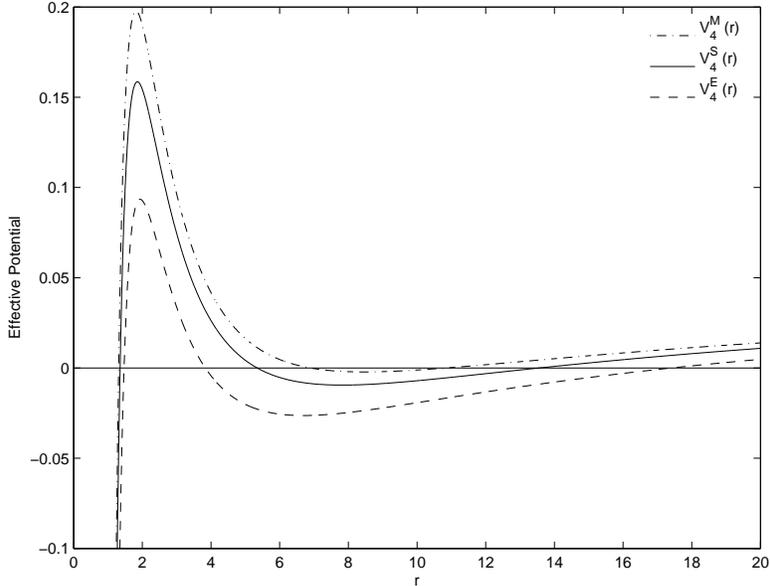}}
\caption{Effective potentials for magnetic  ($V_4^M(r)$, $Q=0.25$), Schwarzschild  ($V_4^S(r)$) and electric ($V_4^E(r)$, $\alpha=0.25$) black holes  with $E=1.95$ and $C=2.2$.}
\label{figp1}
\end{figure}
\begin{figure}
\centerline
\centerline{\includegraphics[scale=0.58]{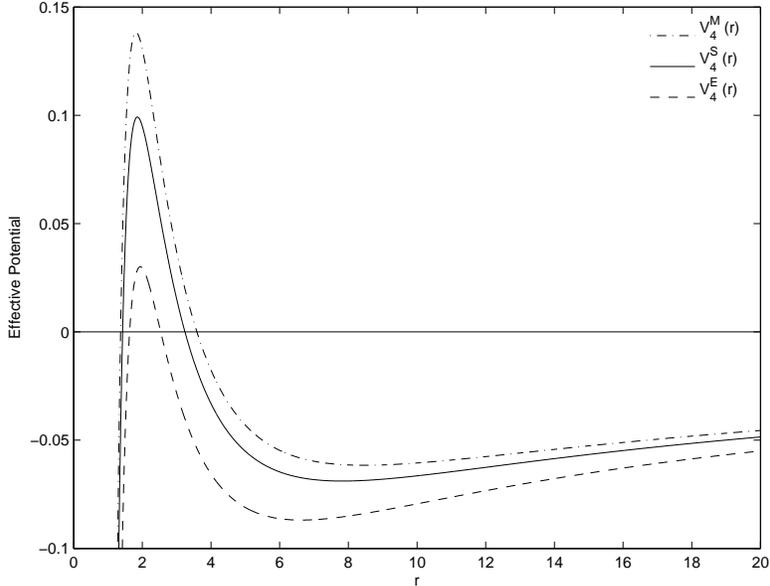}}
\caption{Effective potentials for magnetic  ($V_4^M(r)$, $Q=0.25$), Schwarzschild  ($V_4^S(r)$) and electric ($V_4^E(r)$, $\alpha=0.25$) black holes  with $E=2.01$ and $C=2.2$.}
\label{figp2}
\end{figure}
\noindent Since the Schwarzschild metric reduces to that of the flat spacetime for $r \rightarrow \infty$, the kinematics of deformations far from the singularity is same as for a static flat spacetime, a case which has already been well studied (see \cite{adg}). It may be noticed that the potential (\ref{potentialv4a}) (or (\ref{potentialv4q})) reduces to  (\ref{potsc}) for $\alpha =0$ (or $Q=0$). The effective potentials are graphically presented in Fig. \ref{figp1} (for $E=1.95$) and in Fig. \ref{figp2} (for $E=2.01$).

For the potential in Fig. \ref{figp1}, one can have  bound or infalling trajectories (depending on the value of  $r_0$),  while the potential in  Fig. \ref{figp2} allows infalling and escaping trajectories. Such trajectories are observationally important and therefore, in the following, we study the kinematics of deformations, numerically, in the above--mentioned backgrounds. 

\subsection{Analysis of deformations in the equatorial section \label{v4d}}
In this section, we study the kinematics of deformations restricted to the equatorial section of the 
black hole background.
%Furthermore, to simplify matters, we may choose 
%to study the kinematics of deformations in the following way.
Consider the $\psi=\pi/2$ section of the spacetime which is 
a $2+1$ dimensional slice with an induced metric, say, $\gamma_{\alpha\beta}
$. The timelike geodesic motion is now confined to the $r-\phi$ plane. The geodesics of the induced metric are also geodesics of the
full metric, and the effective potentials remain unaltered. 
We choose the deformations ${\hat \xi}^\alpha$ to be
$2+1$ dimensional and accordingly we define ${\hat B}_{\alpha\beta} =
{\hat \nabla}_\beta {\hat u}_{\alpha}$ where ${\hat u}^{\alpha} \equiv \left (\dot t,\dot r,
\dot \phi \right )$. Similarly, one can also define ${\hat h}_{\alpha\beta} =
\gamma_{\alpha\beta} + {\hat u}_{\alpha} {\hat u}_{\beta}$.  

We then consider a freely falling (Fermi) normal frame $E_{\mu}^\alpha $  (with $E_{t}^\alpha = {\hat u}^{\alpha}$) 
which is parallely transported according to ${\hat u}^{\alpha}{\hat \nabla}_\alpha \, E_{\mu}^{\beta} =0$. 
The kinematics of deformations, restricted to the 2-dimensional spacelike hypersurface (representing the local frame of a
freely falling observer) in this basis, can now be represented by four kinematical quantities, namely,
$\theta$, $\sigma_+$, $\sigma_\times$ and $\omega$. The tensor $\hat B_{\alpha \beta}$ in this basis, can be constructed as,
\begin{equation}
\hat B_{\alpha \beta} = (\frac{1}{2}\theta +\sigma_+) e^r_{\alpha}e^r_{\beta} + (\frac{1}{2}\theta -\sigma_+) e^\phi_{\alpha}e^\phi_{\beta} + (\sigma_\times + \omega)e^r_{\alpha}e^\phi_{\beta} + (\sigma_\times - \omega)e^\phi_{\alpha}e^r_{\beta} \, .\label{bij4d}
\end{equation}
where $e^\mu_{\alpha}$ are co-frame basis satisfying $e^\mu_{\alpha} E_\nu^{\alpha} = \delta^{\mu}_{\nu}$. 
The ESR can be extracted from the evolution tensor (\ref{bij4d}) using the basis vectors as follows, 
\begin{equation}
\theta = \hat B_{\alpha \beta}\, {\hat h}^{\alpha \beta} \equiv \hat B_{\alpha \beta} {\gamma}^{\alpha \beta},
\end{equation}
\begin{equation}
\sigma_+= \frac{1}{2} (\hat B_{\alpha \beta}\,  E_{r}^{\alpha}\, E_{r}^{\beta} - \hat B_{ \alpha \beta} \, E_{\phi}^{\alpha}\, E_{\phi}^{\beta}),
\label{sigf+}
\end{equation} 
\begin{equation}
\sigma_{\times}= \frac{1}{2} ( \hat B_{\alpha \beta} \, E_{r}^{\alpha}\, E_{\phi}^{\beta} + \hat B_{\alpha \beta } \, E_{\phi}^{\alpha}\, E_{r}^{\beta}),
\label{sigfc}
\end{equation} 
\begin{equation}
\omega= \frac{1}{2} ( \hat B_{\alpha \beta} \, E_{r}^{\alpha}\, E_{\phi}^{\beta} - \hat B_{\alpha\beta }\,  E_{\phi}^{\alpha}\, E_{r}^{\beta}).
\label{omf}
\end{equation}

As in the $1+1$ dimensional example discussed earlier, 
the first integrals of the geodesic equations of the $2+1$ dimensional
line element enable us to find the expansion, shear and
rotation for a geodesic congruence. 
Making use of the vector field ${\hat u}^{\alpha}$ and the definition
of ${\hat B}_{\alpha\beta}$, we can obtain $\theta$ for example as follows,
\begin{equation}
\theta = \pm\frac{1}{\dot{r}XY}\left[\frac{E^2}{4r}-\frac{X}{r}-(r^2+C^2)\frac{X'}{2r^2}\right]=
\pm \frac{1}{XY} \frac{\frac{E^2}{4r}-\frac{X}{r} -(r^2+C^2)\frac{X'}{2r^2}}{\sqrt{\frac{E^2}{4XY}-(r^2+C^2)\frac{1}{r^2Y}}},
\end{equation}
where $X(r)$ and $Y(r)$ are the metric functions defined earlier.
Similar general expressions for $\sigma_+$, $\sigma_\times$ and $\omega$ 
can also be obtained. It may be noted that the expressions thus obtained are all functions of $r$. 
Obtaining $r(\lambda)$ by solving the geodesic equations, one can then
find $\theta(\lambda)$, $\sigma_{ij}(\lambda)$ and $\omega_{ij}(\lambda)$. 
From all these expressions, it is easy to state that
divergences in the ESR appear at the turning points (i.e. where $\dot r=0$).
Similarly as in $1+1$ dimensions, these solutions 
represent only special solutions with special initial conditions. 
They do not reveal the effect of initial conditions on the evolution of the ESR variables. 
A more general class of solutions corresponding to arbitrary initial conditions on the ESR variables
are obtained by integrating the full set of Raychaudhuri equations together with the geodesic equations.
%evolve the full set of equations, with given initial conditions, 
%and obtain the $\lambda$ dependence of each kinematical quantity 
%directly.  

In order to understand the caustic formation/focusing behaviour in more detail, let us redefine the expansion scalar as 
$\theta = 2 {\dot F}/F$ where the dot indicates the derivative with respect to $\lambda$. One may then use the 
Fermi normal basis to rewrite (\ref{theta3e}) as the following Hill-type equation,
\begin{equation}
{\ddot F} + H \, F = 0,
\label{theta3e1}
\end{equation}  
where  $ H= \sigma_{+}^2 + \sigma_{\times}^2 -\omega^2 + G$ and $G=R_{\alpha \beta} \, u^{\alpha} \, u^{\beta}/2 $. 

The notion of focusing is related to $F= 0$, $\dot F < 0$  at a finite $\lambda$.  This can be achieved 
under specific conditions on (a) the sign of $H$, 
and (b) the initial values of the ESR. For a complete analysis, we would require to consider all 
possible initial values (positive, negative or zero) for the ESR and $H$. Here, we restrict ourselves to
some special cases and briefly comment on the rest.

\begin{enumerate}
\item[(A)]{If $H>0$ for all $\lambda$, then conjugate points exist and focusing takes place, as is well known.}

\item[(B)]{When $H<0$ for all $\lambda$, we have focusing
only when $\theta_0 < \theta_0^c<0$. When $\theta_0 > \theta_0^c$, there is
defocusing.}

\item[(C)]{If $H$ is sign indefinite over the range of $\lambda$,
then there may be various possibilities depending on the initial conditions of the ESR. 
%If $\theta_0 <0$, then there is focusing.
%Additionally, if $\theta_0>0$, there may exist a critical value $\theta_0^c >0$ (dependent on initial values of
%the shear and rotation), below which we have focusing. 
}
\end{enumerate} 
The conclusion (B) has already been illustrated above for the 2-dimensional case in Section~\ref{2dcase}.

We now illustrate the above conclusions with our numerical evaluations and 
corresponding plots for the ESR variables.
\begin{figure}
%\begin{center}
\begin{picture}(300,270)(0,0)
\put(-60,-20){
{\includegraphics[scale=0.9]{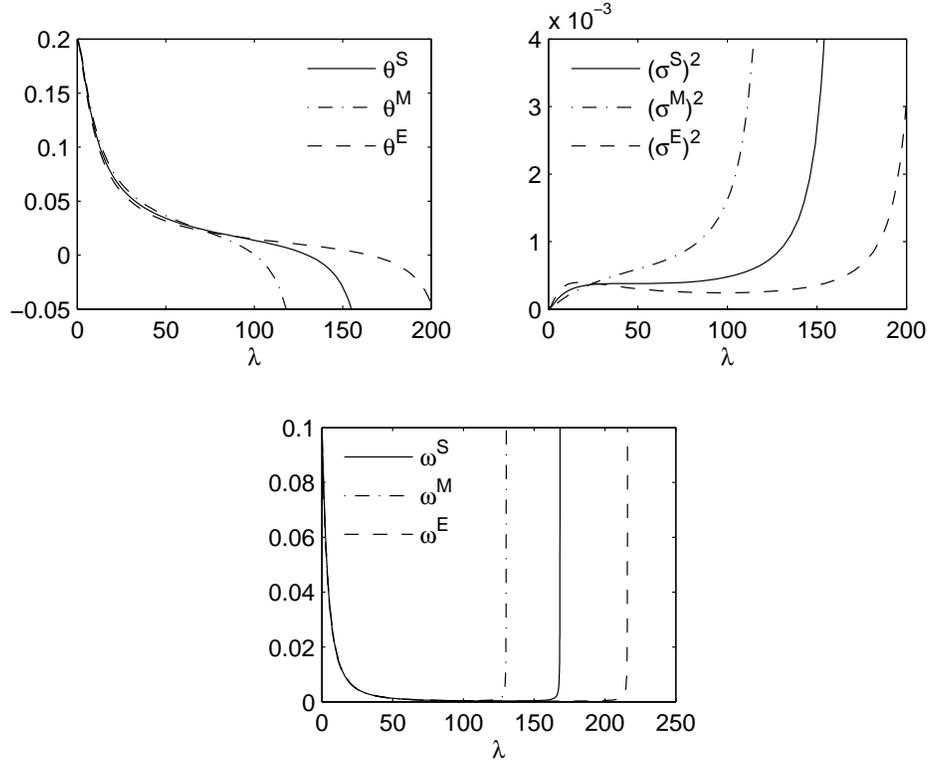}}
}
\end{picture}
%\end{center}
\caption{Comparison of ESR variables in electric ($\alpha= 0.25$), magnetic ($Q=0.25$) and Schwarzschild black holes with $E=1.95$, $C=2.2$, $\theta_0 = 0.2$, $\sigma_{+0} = 0$, $\sigma_{\times 0} = 0$, $\omega_0 =0.1$ and $r_0 =8$.}
\label{fig1cmp}
\end{figure}
\begin{figure}
\begin{center}
{\includegraphics[scale=0.7]{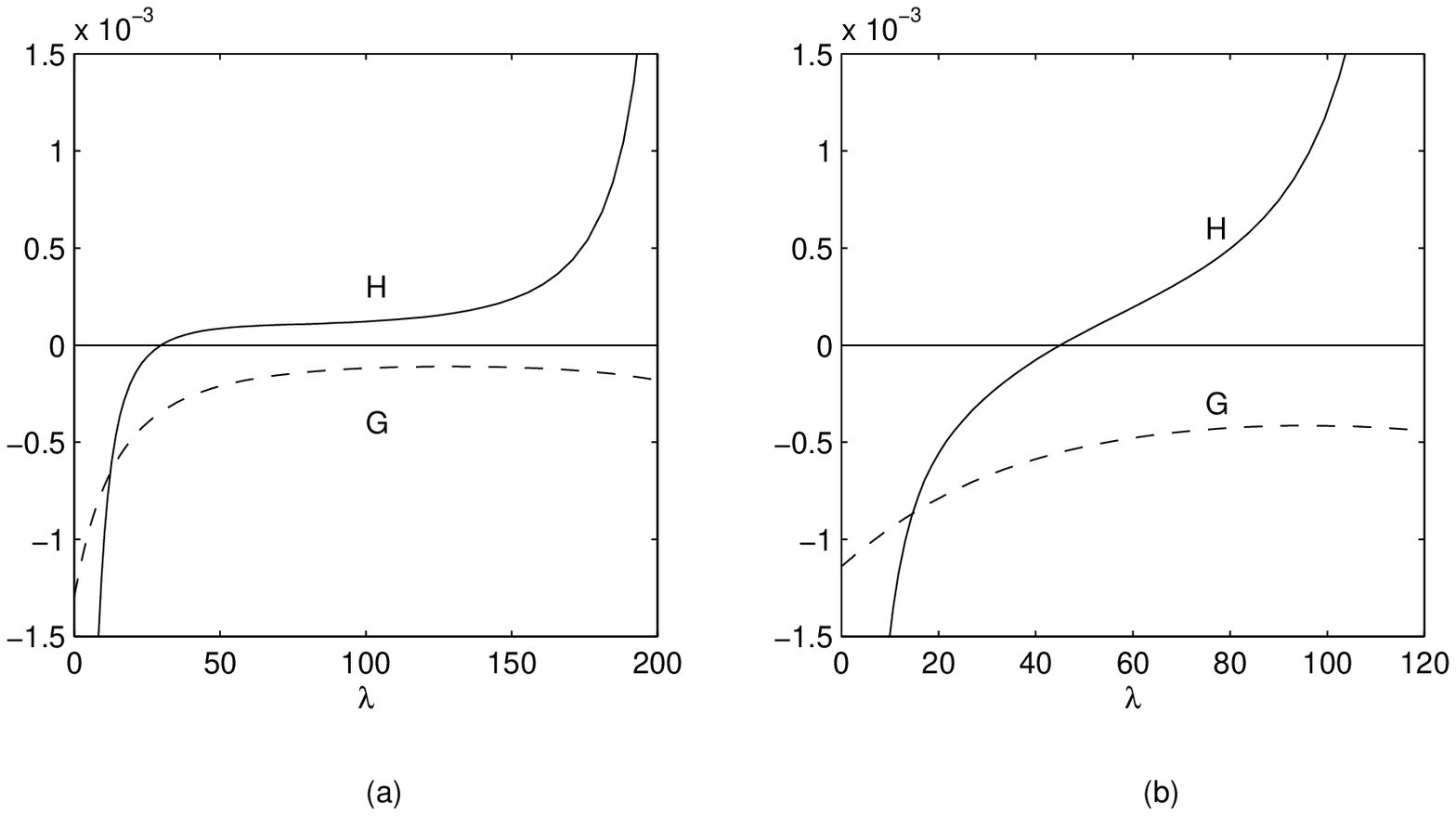}}
\end{center}
\caption{ Variation of $G$ and $H$ for (a)  electric and, (b) magnetic black holes with the initial conditions as for Fig.~\ref{fig1cmp}.}
\label{fig2cmp1}
\end{figure}
%\begin{figure}
%\centerline
%\centerline{\includegraphics[scale=0.58]{schemfig.eps}}
%\caption{Schematic diagram for explaining the occurrence of finite time singularity in the expansion scalar.}
%\label{figsch}
%\end{figure}
\begin{figure}
\centerline
\centerline{\includegraphics[scale=0.58]{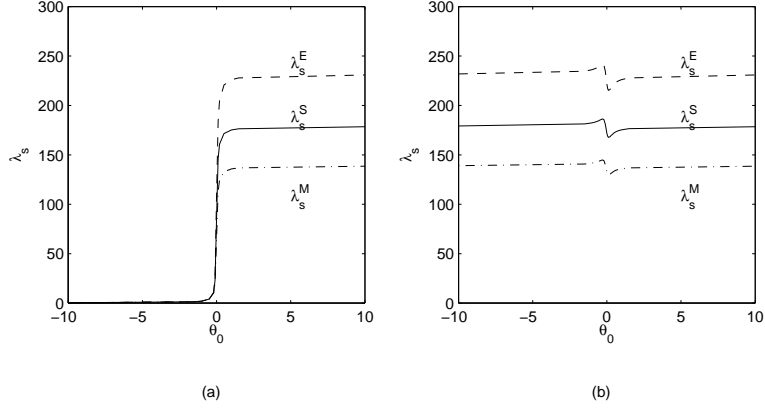}}
\caption{Variation of time to singularity $(\lambda_s)$ versus initial expansion $(\theta_0 )$ for different black hole metrics (a) without initial rotation of the congruence, and (b) with initial rotation of the congruence $(\omega_0 = 0.1$). Here, we have considered $\sigma_{+0}=0$ and $\sigma_{\times 0}=0$.}
\label{figtos}
\end{figure}
\begin{figure}
%\begin{center}
\begin{picture}(300,270)(0,0)
\put(-60,-20){
{\includegraphics[scale=0.9]{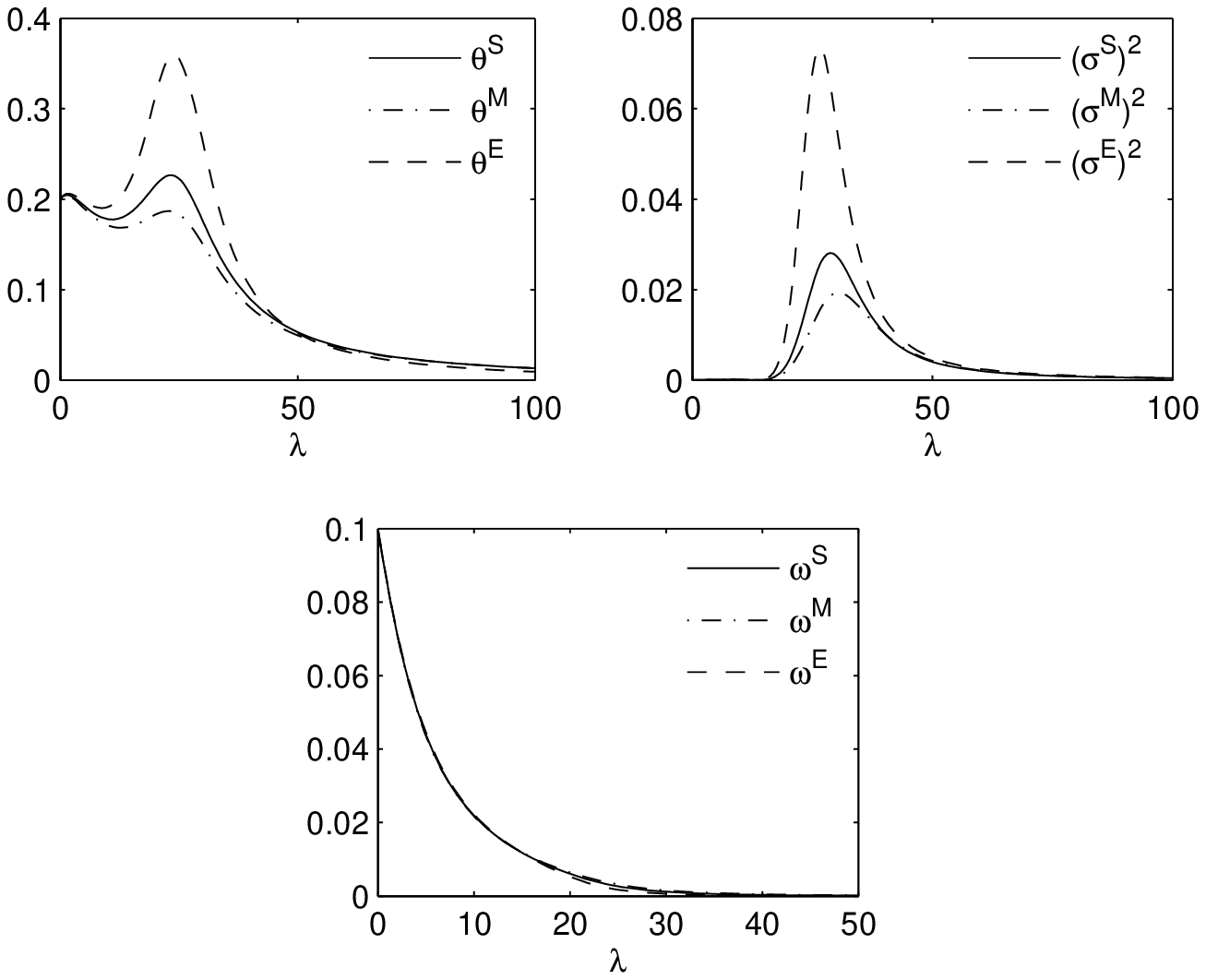}}
}
\end{picture}
%\end{center}
\caption{Comparison of ESR variables in electric ($\alpha= 0.25$), magnetic ($Q=0.25$) and Schwarzschild black holes with $E=2.01$, $C=2.2$, $\theta_0 = 0.2$, $\sigma_{+0} = 0$, $\sigma_{\times 0} = 0$, $\omega_0 =0.1$ and $r_0 =8$.}
\label{fig2cmp}
\end{figure}
\begin{figure}
\begin{center}
{\includegraphics[scale=0.7]{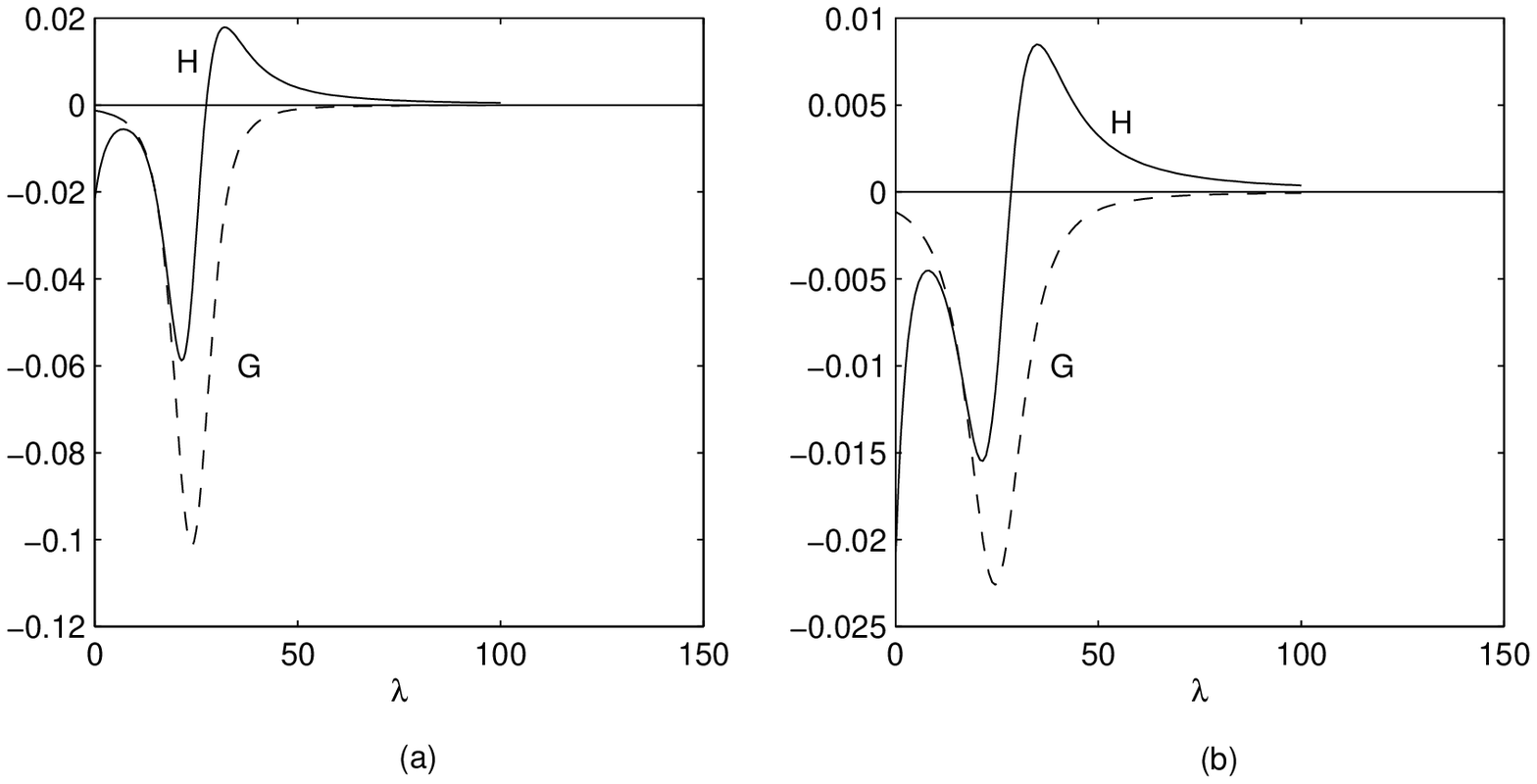}}
\end{center}
\caption{Variation of $G$ and $H$ for (a)  electric and, (b) magnetic black holes with the initial conditions as for Fig.~\ref{fig2cmp}.}
\label{fig2cmp2}
\end{figure}
In what follows, we analyse and compare the kinematics of deformations in the charged (electric and magnetic) stringy black hole with the Schwarzschild black hole. For this, we consider the two potentials shown in Figs.~\ref{figp1} and \ref{figp2} for the bound and escaping trajectories for the reason mentioned before. Corresponding to these two potentials, the ESR variables for the two black holes are compared below. In both the cases, the initial radius is taken as $r_0=8$.
The initial $\dot r$ is considered as positive for the evaluations for
 Fig.~\ref{fig1cmp}, while $\dot r$ is taken as negative for the evaluations depicted
in Fig.~\ref{fig2cmp}.
\\
In Fig.~\ref{fig1cmp} (corresponding to the the potential shown in Fig.~\ref{figp1}), we observe geodesic focusing in all the three backgrounds. 
In order to understand this behaviour in light of the general conclusions mentioned earlier, we have plotted $H$ and $G$
in Fig. 4, for geodesic congruences in the electric and magnetic stringy black holes. It is evident 
that $G$ is negative definite along the geodesic flow.  
However, $H$ is initially negative over a certain range of $\lambda$ but 
soon turns positive and
increases monotonically due to dominating contributions from the shear components. Thus, one may say that the focusing effect seen here is largely 
shear--induced. Even
though the rotation term appears with an opposite sign in $H(\lambda)$, the net
value, beyond a certain $\lambda$ becomes positive and may be responsible for
the focusing effect. We may note, in passing, that the focusing effect in 
Schwarzschild spacetime is entirely due the eventual dominance 
of shear over rotation. 
%In all cases, we have found that with the variation of the initial rotation, the focusing effect gets modified. 
%Hence, the conclusions in the case (B) (with $0<\theta_0<\theta_0^c$) discussed above apply for which we have focusing.

%In order to understand this focusing behaviour in more details, let us redefine the expansion scalar as $\theta = 2 {\dot F}/F$ where the dot indicates the derivative with respect to $\lambda$. One may then use the Fermi normal basis to rewrite (\ref{theta3e}) as the following Hill-type equation,
%\begin{equation}
%{\ddot F} + H \, F = 0,
%\label{theta3e1}
%\end{equation}  
%where  $ H= \sigma_{+}^2 + \sigma_{\times}^2 -\omega^2 + G$ and $G=R_{\alpha \beta} \, u^{\alpha} \, u^{\beta}/2 $. It is clear from (\ref{theta3e1}) that if $\ddot F <0$ for  $F>0$ and vice-versa. Therefore, $F$ will eventually go to zero in finite time for arbitrary initial condition $\theta_0$ . This would imply that, under this condition, $\theta \to -\infty$ in finite time even for an  initially expanding congruence.
The time of approach to a singularity in the congruence, which we denote by $\lambda_s$, can also be
an interesting quantifier which we can use to characterise geodesic flows in the three backgrounds. 
We have numerically studied the effect of the initial expansion $\theta_0$ on  $\lambda_s$  for the different black hole metrics without and with initial rotation $\omega_0$ of the congruence. The results of this study are shown in Fig. \ref{figtos}. It is interesting to note that for an initially contracting congruence the singularity occurs more rapidly as compared to an initially expanding congruence. Furthermore, the time to singularity does not change appreciably for large values of initial expansion/contraction. On the other hand, even with a small initial rotation ($\omega_0 =0.1$), $\lambda_s$ remains almost unchanged over the whole range of variation of the $\theta_0$ considered. It was also found (though the results are not presented) that, with initial shear, $\lambda_s$ reduces drastically over the complete range of $\theta_0$ which is also expected qualitatively (see \cite{review}). One difference that emerges from this study on the three black hole metrics considered is that $ \lambda_s^E > \lambda_s^S > \lambda_s^M $ as observed in Fig. \ref{figtos}.

In Fig.~\ref{fig2cmp} (corresponding to the the potential shown in Fig.~\ref{figp2}), there is no focusing. 
We observe from Fig.~\ref{fig2cmp2} that $H$ is sign indefinite and bounded over the range of $\lambda$. As mentioned
in (C) above, in such situations, definite conclusions are difficult to arrive at because of the sensitive dependence of the results on the 
initial values of the ESR. The following curious features may be associated with the no--focusing behaviour observed above. Firstly,
it is easy to note from the Fig.~\ref{fig2cmp2} that $\int H \,d\lambda $ has a negative value, unlike
that observed in Fig. 4. Further, the nature of $H$ (and $\dot H$) in 
Figs. 4 and \ref{fig2cmp2}, for large $\lambda$, are distinctly different
and may also be a cause of the no-focusing effect.

Additionally, the location (in $\lambda$) of the extrema (maxima/minima) in the ESR
%$\theta$, $\sigma_+$ and $\sigma_\times$ 
shifts to larger $\lambda$ as we move from the
electric to Schwarzschild to the magnetic solutions. This is evident in Fig. 6.
If $\lambda_e$ denotes the location of an extremum, from the figure, we
can easily say that $\lambda_e^E>\lambda_e^S> \lambda_e^M$.  
 
%Moreover, it is also possible that the rate of fall off $H$ in
Finally, it may be noted from the plots in Fig. 6 corresponding to $\omega$ that the rotation of the congruence is 
largely similar irrespective of the metric. This is expected since $\omega = \omega_0 \,\exp( -\int \theta \, d t) $, 
and the variation of $\theta$ is almost similar in the three backgrounds. In contrast, from this expression
of $\omega$, we note that
the divergence (to $-\infty$) of $\theta$ is reflected in the divergence (to $+\infty$) 
of $\omega$, as shown in Fig.~\ref{fig1cmp}.

\section{Summary and Conclusions}
In this article, we have investigated the kinematics  
of timelike geodesic flows in two and four dimensional spacetimes 
representing stringy black holes. 
We now briefly summarise the work done, the conclusions 
drawn from it and also mention possibilities on future work.
\begin{itemize}
\item [(i)]{The exact solutions of the expansion scalar for the different cases in 
2D have been calculated by solving the corresponding Raychaudhuri equation
for the expansion. The occurrence of a finite time singularity 
(i.e., caustic formation/geodesic focusing) in each case is then discussed with
particular reference to the relation between initial conditions and the behaviour of the expansion.} 
\item [(ii)] {The geodesic equations and the Raychaudhuri equations for the
ESR are written out and solved 
numerically for timelike geodesic congruences in two different stringy black hole spacetimes in four dimensions.  }

\item [(iii)] {We have drawn some general conclusions and made some observations on geodesic focusing
which we believe are new. In particular, we have demonstrated how different initial conditions on the ESR can affect the
occurence of geodesic focusing. We also show how focusing can be affected by the variation of $H(\lambda)$.
Even in situations where the timelike convergence condition is violated, domination of shear can still
lead to focusing. Further, we have introduced a new quantity - the time of approach to singularity,
which may be used to distinguish between geodesic flows in different backgrounds.
Though not presented in this article, we have observed that, in the presence of initial shear 
(i.e., $\sigma_{+0}\neq 0$ and/or $\sigma_{\times 0}\neq 0$), 
the time of approach to singularity ($\lambda_s$) is significantly reduced. }

%In all the cases, even with zero-valued initial conditions, the evolved 
%values of the ESR variables get highly enhanced  near the horizon 
%because of the effect of a larger curvature. However, far from the horizon the ESR variables are less enhanced due to relatively low value of the curvature and qualitatively, their evolution is nearly similar as in the case of a flat space.}

\item [(iv)] {In the scenario depicted in Fig. 6, where there is no focusing,
we make an attempt towards understanding why this happens by analysing
the behaviour. We observe here that the locations of the extrema
show a systematic shift as we move from geodesic flows in the
electric to the Schwarzschild and then to the magnetic black holes.}

\item [(v)] {On the whole, in some sense, the stringy nature of the black hole
geometry does seem to manifest itself in the nature of evolution of the ESR.}

\end{itemize}

An interesting issue that is still left unanswered in this work in the role
of duality of the electric and magnetic black hole metrics on
the kinematics. The question that might be asked is whether the
kinematics in these two spacetimes are also dual of one another in
some sense. It may be tempting to approach this issue by searching 
relations between the parameters $\alpha$, $Q$, $m$, $E$ and $C$ which
leaves the kinematics invariant.

The metrics of the stringy black holes have coordinate singularities
at specific values of $r$ and hence cannot be extended beyond. 
Thus, for a more complete description of the kinematics of flows, it will be 
interesting to study geodesic flows using a different, extendable
coordinate system (viz. the maximally extended  Kruskal coordinate system). 
%Moreover, the detailed analysis of different kinds of possible trajectories 
%based on various conditions on constants of motion will also be investigated in our forthcoming studies. 
Besides this, our work has so far focused entirely on timelike geodesics in static spacetimes. It would be worth
studying the nature of null congruences in a similar fashion. A logical next step would be to consider geodesic 
flows in stationary metrics (such as the Kerr black hole and its generalisations).
 
Finally, the essential goal behind this work has been to demonstrate 
that the Raychaudhuri equations and the geodesic equations can be
solved simultaneously to give us a complete picture of geodesics and
geodesic flows in any given spacetime. Thus, we have a viable approach for
studying the kinematics of geodesic congruences for any given metric which can help us distinguish between spacetimes through the
behaviour of trajectories and families of trajectories. 
In the long run, it may be possible to make use of these results
in arriving at distinct observable effects in specific gravitational fields.\\

\section*{Acknowledgments}
\noindent The authors thank the Department of Science and Technology (DST), Government of India for financial support through  
a sponsored project (Grant Number: SR/S2/HEP-10/2005). The authors also thank the anonymous referee for his constructive
comments which helped in improving the presentation of the paper.
%One of the authors (HN) would also like to thank 
%Drs Ratna Koley  and Suprtaik Pal for some helpful discussions during the early stage of this work.


\begin{references}
\bibitem{toolkit} E. Poisson, {\em A relativists' toolkit: the mathematics
of black hole mechanics} (Cambridge University Press, UK, 2004).
\bibitem{wald} R. M. Wald, {\em General Relativity} (University of Chicago
Press, Chicago, USA, 1984).
\bibitem{joshi} P. S. Joshi, {\em Global aspects in gravitation and 
cosmology} (Oxford University Press, Oxford, UK, 1997).
\bibitem{ellis} G. F. R. Ellis in {\em General Relativity and Cosmology},
{\em International School of Physics, Enrico Fermi--Course XLVII} (Academic
Press, New York, 1971).
\bibitem{ciufolani} I. Ciufolini and J. A. Wheeler, {\em Gravitation and
inertia} (Princeton University Press, Princeton, USA, 1995).
\bibitem{review} S. Kar and S. SenGupta, Pramana {\bf 69}, 49 (2007); 
gr-qc/0611123 and references therein; S. Kar, {\em Introducing the Raychaudhuri equations}, Resonance, Journal of Science Education {\bf 13}, 319 (2008).
\bibitem{hawk} S. W. Hawking and G. F. R. Ellis, {\em The large scale
structure of spacetime} (Cambridge University Press, Cambridge, UK, 1973).
\bibitem{Penro} R. Penrose, {\em Gravitational collapse and space-time 
singularities}, Phys. Rev. Lett. {\bf 14}, 57 (1965).
\bibitem{Haw} S. W. Hawking,{\em Occurrence of singularities in open 
universes}, Phys. Rev. Lett. {\bf 15}, 689 (1965); {\em Singularities in the 
universe},ibid {\bf 17}, 444 (1966).
\bibitem{adg1} A. Dasgupta, H. Nandan and S. Kar, {\em Kinematics of deformable media}, Annals of Physics {\bf 323}, 1621 (2008); \, arXiv : 0709.0582.
\bibitem{adg} A. Dasgupta, H. Nandan and S. Kar, {\em Kinematics of flows on 
curved, deformable media}, arXiv : 0804.4089 [Int. J. of
Geom. Meth. Mod. Phys.  {\bf 6(4)} (2009) in press].
\bibitem{prd} F. Shojai, A. Shojai, {\em Geodesic Congruences in the Palatini f(R) Theory}, Phys. Rev. {\bf D78}, 104011 (2008).
\bibitem{Witten} E. Witten, {\em String theory and black holes}, Phys. Rev. {\bf D44}, 314 (1991). 
\bibitem{Sen} A. Sen, {\em Rotating charged black hole solution in heterotic string theory}, Phys. Rev. Letts. {\bf 69}, 1006 (1992). 
\bibitem{grumiller} D. Grumiller, W. Kummer and D. V. Vassilevich, {\em Dilaton gravity in two dimensions}, Phys. Rep. {\bf 369}, 327 (2002). 
\bibitem{harvey} J. Harvey and A. Strominger, {\em Quantum aspects of black holes in string theory and quantum gravity}, in String Theory and Quantum Gravity 92, edited by J Harvey et al. (World Scientific, Singapore, 1993).
\bibitem{Mandal} G. Mandal, A. M. Sengupta and S. R. Wadia, {\em Classical solutions of 2-dimensional string theory}, Mod. Phys. Letts. {\bf A6}, 1685 (1991).
\bibitem{rk} R. Koley, S. Pal and S. Kar, {\em Geodesics and geodesic deviation in a two-dimensional black hole},  Am. J. Phys. {\bf 71}, 1037 (2003).
\bibitem{ghs} D. Garfinkle, G. T. Horowitz and A. Strominger, {\em Charged black holes in string theory},
Phys. Rev. {\bf D43}, 3140 (1991), {\em Erratum ibid} Phys. Rev.{\bf D45}, 
3888 (1992).
\bibitem{gthorowitz} G. T. Horowitz, {\em The dark side of string theory : black holes and black strings}, hep-th/9210119.
\bibitem{gsw} M. S. Green, J. H. Schwarz and E. Witten, {\em
Superstring theory} (Cambridge University Press, UK, 1987); J. Polchinski,
{\em String theory} (Cambridge University Press, UK, 1997).
\bibitem{skprd97} S. Kar,  {\em Stringy black holes and energy conditions},
  Phys. Rev. {\bf D55}, 4872 (1997) and references therein. 
\bibitem{tipler} F. J. Tipler, {\em Energy conditions and spacetime
    singularities}, Phys. Rev.{\bf D17},        
 2521(1978); F. J. Tiper, {\em Singularities and Causality Violation},
 Ann. Phys. (N.Y.) {\bf 108}, 1 (1977).
\bibitem{Hartle} J. M. Hartle, {\em Gravity  An Introduction to Einstein's General Relativity} (Pearson Education Inc., Singapore 2003).

\end{references}
\end{document}